\documentclass[twocolumn,amsmath,aps,fleqn]{revtex4}
\usepackage{graphicx,amssymb}
\begin{document}
\newcommand{\be}{\begin{equation}}
\newcommand{\ee}{\end{equation}}
\newcommand{\bq}{\begin{eqnarray}}
\newcommand{\eq}{\end{eqnarray}}
\newcommand{\bsq}{\begin{subequations}}
\newcommand{\esq}{\end{subequations}}
\newcommand{\bc}{\begin{center}}
\newcommand{\ec}{\end{center}}
\newcommand {\R}{{\mathcal R}}
\newcommand{\al}{\alpha}
\newcommand\lsim{\mathrel{\rlap{\lower4pt\hbox{\hskip1pt$\sim$}}
    \raise1pt\hbox{$<$}}}
\newcommand\gsim{\mathrel{\rlap{\lower4pt\hbox{\hskip1pt$\sim$}}
    \raise1pt\hbox{$>$}}}

\title{Thermodynamics of Cosmic Defect Network Evolution}

\author{P.P. Avelino}
\email[Electronic address: ]{pedro.avelino@astro.up.pt}
\affiliation{Instituto de Astrof\'{\i}sica e Ci\^encias do Espa{\c c}o, Universidade do Porto, CAUP, Rua das Estrelas, PT4150-762 Porto, Portugal}
\affiliation{Centro de Astrof\'{\i}sica da Universidade do Porto, Rua das Estrelas, PT4150-762 Porto, Portugal}
\affiliation{Departamento de F\'{\i}sica e Astronomia, Faculdade de Ci\^encias, Universidade do Porto, Rua do Campo Alegre 687, PT4169-007 Porto, Portugal}

\author{L. Sousa}
\email[Electronic address: ]{Lara.Sousa@astro.up.pt}
\affiliation{Instituto de Astrof\'{\i}sica e Ci\^encias do Espa{\c c}o, Universidade do Porto, CAUP, Rua das Estrelas, PT4150-762 Porto, Portugal}
\affiliation{Centro de Astrof\'{\i}sica da Universidade do Porto, Rua das Estrelas, PT4150-762 Porto, Portugal}

\date{\today}
\begin{abstract}

We show that simple thermodynamic conditions determine, to a great extent, the equation of state and dynamics of cosmic defects of arbitrary dimensionality. We use these conditions to provide a more direct derivation of the Velocity-dependent One-Scale (VOS) model for the macroscopic dynamics of topological defects of arbitrary dimensionality in a $N+1$-dimensional homogeneous and isotropic universe. We parameterize the modifications to the VOS model associated to the interaction of the topological defects with other fields, including, in particular, a new dynamical degree of freedom associated to the variation of the mass per unit $p$-area of the defects, and compute the corresponding scaling solutions. The observational impact of this new dynamical degree of freedom is also briefly discussed.

\end{abstract}
\maketitle

\section{\label{intr}Introduction}

The production of topological defects as remnants of phase transitions in the early universe is expected in many grand unified scenarios and string-theory inspired inflationary models. In these models, a significant contribution to the gravitational field --- albeit subdominant, at least on large cosmological scales --- is causally seeded by defect evolution on increasingly larger scales. Topological defects may then generate specific signatures on a wide range of cosmological observations. These include the cosmic microwave background temperature and polarization anisotropies \cite{Pogosian:1999np,Pogosian:2007gi,Ade:2013xla,Lazanu:2014xxa,Lizarraga:2014xza,Sousa:2015cqa}, the stochastic gravitational wave background \cite{Damour:2004kw,Olmez:2010bi,Hiramatsu:2010yz,Sanidas:2012ee,Binetruy:2012ze,Hiramatsu:2013qaa,Sousa:2013aaa,Sousa:2014gka}, small-scale structure formation \cite{Wu:1998mr,Pogosian:2008am,Lin:2015asa,Anthonisen:2015tda,Bramberger:2015kua} and reionization \cite{Avelino:2003nn,Avelino:2004wm,Pogosian:2004mi,Olum:2006at}, and gravitational lensing observations \cite{deLaix:1997jt,Mack:2007ae,Thomas:2009bm,Yamauchi:2011cu}. A quantitative forecast of the distinct observational signatures associated to cosmic defects requires a detailed understanding of the evolution of these networks. Although this may be achieved, for specific models, using high-resolution numerical simulations, semi-analytical multi-parameter frameworks are often more flexible and informative.

Initial attempts to model the macroscopic dynamics of defect networks \cite{Kibble:1984hp,Bennett:1985qt,Bennett:1986zn} were focused on cosmic strings and assumed that a single dynamical variable --- the characteristic length of the network --- is sufficient to describe their dynamics. A more sophisticated macroscopic model, the Velocity-dependent One-Scale (VOS) model, which incorporates the root-mean-squared (RMS) velocity of the cosmic string network as a dynamical variable, was later proposed in \cite{Martins:1996jp,Martins:2000cs}. This model has been thoroughly tested and calibrated using numerical simulations and later extended to account for the dynamics of domain walls \cite{Avelino:2005kn,Avelino:2011ev}. More recently, a  unified framework for the macroscopic evolution of featureless $p$-branes of arbitrary dimensionality in $N+1$-dimensional homogeneous and isotropic spacetimes (with $N>p$) was rigorously derived in \cite{Sousa:2011ew,Sousa:2011iu}. This unified paradigm provides a common tool for describing defect dynamics in cosmology as well as in a wide variety of other contexts (including condensed matter \cite{Avelino:2010qf} and biology \cite{2012PhRvE..86c1119A}).

In this paper, we provide a unified thermodynamic derivation of the VOS equations of motion for the macroscopic dynamics of cosmic defects, including the defect mass per unit $p$-dimensional area as a new dynamical degree of freedom. In Sec. \ref{sec2}, we start by showing that the VOS model for the evolution of a network of point particles, interacting minimally with each other and with other fields, may be derived using simple thermodynamic conditions. In Sec. \ref{sec3}, we extend the results of the previous section to describe the dynamics of non-interacting featureless planar domain wall networks in $3+1$ dimensions, while assuming entropy conservation in a homogeneous universe. In Sec. \ref{sec4}, we further generalize our results for a network of topological defects of arbitrary dimensionality in a $N+1$-dimensional homogeneous and isotropic universe, incorporating the effect of friction and of a new dynamical variable associated to the variation of the defect mass per unit $p$-dimensional area. After incorporating in the model the conventional curvature and energy loss parameters, we compute the corresponding scaling solutions. We then conclude in Sec.  \ref{conc}.

Throughout this paper we use units such that $c=1$, where $c$ is the value of the speed of light in vacuum.

\section{The simplest case: network of point particles in $3+1$ dimensions \label{sec2}}

Let us start by considering the case of point particles (or $0$-branes) in a $3+1$-dimensional Friedmann-Lemaitre-Robertson-Walker (FLRW) universe, whose line element is given by

\be
ds^2=dt^2-a^2(t)d\mathbf{x} \ldotp d \mathbf{x}\,,
\label{FLRW}
\ee
where $a$ is the cosmological scale factor, $t$ is the physical time and $\bf{x}$ is a $3$-dimensional vector whose components are comoving coordinates. Note that Eq. (\ref{FLRW}) also applies to FLRW universes of arbitrary dimensionality $N+1$. However, in that case, $\bf{x}$ is a $N$-dimensional vector.

\subsection{Linear momentum evolution: Thermodynamics}

Consider a statistically homogeneous and isotropic network of point particles moving with the same speed $v$, while interacting minimally with each other and with all other fields. Each particle then has an energy
\be
E= m \gamma\,,
\ee
where $\gamma \equiv (1-v^2)^{-1/2}$ is the Lorentz factor and $m \equiv m(v=0)$ is the rest mass of the particles. If one assumes that the number of particles and particle mass are conserved, the average particle energy density scales as 
\be
{\bar \rho} \propto \gamma a^{-3}\,.
\label{density0}
\ee
 The average pressure of the network is given by
\be
{\cal {\bar P}}=\frac13 {\bar \rho} v^2\,.
\label{pressure0}
\ee
In a homogeneous and isotropic universe, which evolves adiabatically, the first law of thermodynamics implies that
\be
dU + {\cal {\bar P}} dV = 0\,,
\ee
where $U \equiv {\bar \rho} V$ is the internal energy and $V \propto a^{3}$ is the physical volume. Equivalently, this implies that
\be
{\dot {\bar \rho}} + 3H({\bar \rho} + {\cal {\bar P}})=0\,,
\label{entropyc}
\ee
where $H={\dot a}/a$ is the Hubble parameter and a dot represents a derivative with respect to the physical time. One then finds, by substituting Eq. (\ref{pressure0}) into Eq. (\ref{entropyc}), that the energy density of a network of point particles evolves as
\be
\dot {\bar \rho}+ H {\bar \rho}(3 + v^2)=0\,.
\label{rhoeq}
\ee
This equation, together with Eq. (\ref{density0}), allows us to obtain the following equation of motion for the velocity of the particles
\be
{\dot v} + H(1-v^2)v=0\,,
\label{veq}
\ee
which implies that the linear momentum of the particles is conserved:
\be
\gamma v \propto a^{-1}\,.
\ee
The evolution of the linear-momentum of a free particle in a FLRW universe is, therefore, a direct consequence of the adiabatic evolution of the background and of the first law of thermodynamics.

\subsection{\label{kin}Linear momentum evolution: Kinematics}

The evolution of the  linear-momentum of free particles in a homogeneous and isotropic universe is also a result of the kinematic effect associated to the relative velocity between comoving observers. Consider two comoving observers $O$ and $O'$ and assume that $O'$ is moving with respect to $O$ with relative speed $ \delta u$ along the positive $z$-direction. Let us now consider a particle moving along the positive $z$-direction with a speed $v$ relative to $O$. The velocity of this particle as measured by the observer $O'$ is given by the relativistic velocity-addition formula
\be
v'=\frac{v -\delta u}{1-v \delta u}\,.
\label{vprime1}
\ee
If the two observers are infinitesimally close to each other, then
\be
\delta u=d u = H dr = H v dt \,,
\label{deltau}
\ee
where $dr$ is the physical distance between the two observers at the time of the velocity  measurements by the observers $O$ and $O'$ (made at the instants $t$ and $t+dt$, respectively). Substituting Eq. (\ref{deltau}) into Eq. (\ref{vprime1}), one obtains
\be
v'=\frac{v -Hvdt}{1-H v^2 dt} = v + H v (v^2-1) dt\,,
\label{vprime}
\ee
up to first order in $dt$, which is equivalent to Eq. (\ref{veq}) (note that $du=v'-v$).

\subsection{\label{vos} VOS model}

Let us now relax the assumption that all particles have the same speed, and assume that the network is statistically homogeneous and isotropic on large scales. In this case, one may obtain an evolution equation for the RMS velocity of the network $\bar v$, by  multiplying Eq. (\ref{veq}) by $v$ and performing an average. One then has
\be
{\dot {\bar v}} + H(1-{\bar v}^2){\bar v}=0\,,
\label{veqvos}
\ee
were, ${\bar v}^2 \equiv \langle v^2 \rangle$, with
\be
\langle v^2 \rangle \equiv \frac{\sum v^2_i E_i}{\sum E_i}\,,
\ee
and the sum is performed over a very large number of particles $N$ with different energies $E_i$ and speeds $v_i$ in a large comoving volume of the universe. Note that we have assumed, in the derivation of Eq. (\ref{veqvos}), that $\langle v^4 \rangle={\bar v}^4$. Although this relation is true in the ultra-relativistic limit, in other regimes it underestimates the value of  $\langle v^4 \rangle$ since
\be
\langle v^4 \rangle = {\bar v}^4 +\langle (v^2-{\bar v}^2)^2 \rangle\,.
\ee
For the extreme case of a uniform distribution of $v^2$ in the interval $[0,2{\bar v}^2]$, it is straightforward to show that, in the non-relativistic regime,
\be
\langle (v^2-{\bar v}^2)^2 \rangle=\frac13 {\bar v}^4\,,
\ee
and, therefore, this approximation is always fairly good. Note also that in this regime $\langle v^4 \rangle$ is necessarily very small.
 
Moreover, one may obtain an evolution equation for the characteristic length of the network $L$ --- usually defined as
\be
{\bar \rho}=\frac{m}{L^3}\,,
\label{Ldef}
\ee
--- by averaging Eq. (\ref{rhoeq}).  This yields
\be
{\dot L}-\left(1+\frac{{\bar v}^2}{3}\right)HL=0\,.
\label{Leqvos}
\ee
Eqs. (\ref{veqvos}) and (\ref{Leqvos}) constitute the VOS model for a statistically homogeneous and isotropic network of free point particles in a $3+1$-dimensional FLRW universe.

\section{Domain Wall network in $3+1$ dimensions \label{sec3}}

In this section, we extend the results of the previous section for networks of non-interacting featureless planar domain walls ($2$-branes) in $3+1$ dimensions, again using simple thermodynamic conditions associated to energy conservation in a homogeneous and isotropic universe.

\subsection{Energy-momentum tensor}

Consider a static planar domain wall oriented along the $z$-direction. If one assumes that the domain wall is featureless, the physical velocity must be perpendicular to the plane of the wall. Consequently, the components of the energy-momentum tensor must be invariant under Lorentz boosts in any direction along the wall (we shall assume that the local perturbations to the FLRW geometry due to the domain walls can be neglected).

Under a Lorentz  boost in the $x$-direction --- defined by
\bq
\Lambda^{0'}_0&=&\Lambda^{x'}_x=\gamma \,, \quad \Lambda^{0'}_x=\Lambda^{x'}_0=\gamma v \,,\\
\Lambda^{y'}_y&=&\Lambda^{z'}_z=1\,,
\eq
with all other components of $\Lambda^{\mu'}_{\nu}$ vanishing ---, the components of the energy-momentum tensor transform as
\be
T^{\mu' \nu'}=\Lambda^{\mu'}_{\alpha} \Lambda^{\nu'}_{\beta}  T^{\alpha \beta}\,.
\ee
Imposing that $T^{\mu' \nu'}=T^{\mu \nu}$, one finds that 
\bq
T^{0x}&=&T^{0y}=T^{0z}=T^{xy}=T^{xz}=0 \,, \label{Tboost1}\\
\rho &\equiv& T^{00}=-T^{xx}\,. \label{Tboost2}
\eq
By performing a Lorentz boost along the $y$-direction, one obtains similar conditions to Eqs. (\ref{Tboost1}) and (\ref{Tboost2}) with $x$ replaced by $y$. Given the fact that the energy-momentum tensor is symmetric, the only new conditions that emerge are $T^{yz}=0$ and $\rho \equiv T^{00}=-T^{yy}$. These conditions imply that all off-diagonal  components of the energy-momentum tensor of a static domain wall vanish.

Let us now consider a statistically homogeneous (but anisotropic) static planar domain wall network wherein domain walls are oriented along the same direction (say the $z$-direction). Let us also assume an anisotropic expansion along the $x$-direction (or $y$-direction) only, so that the physical volume grows as $V \propto a$. The first law of thermodynamics implies that
\be
dU + {\cal {\bar P}}_\parallel dV=0\,,
\label{entropyeq1}
\ee
where ${\cal {\bar P}}_\parallel$ is the weighted volume average of $T^{xx}$ (or $T^{yy}$),
\be
{\cal {\bar P}}_\parallel=\frac{\int d V T^{xx}}{\int d V}=\frac{\int d V T^{yy}}{\int d V}\,,
\ee
and $U={\bar \rho} V$ with
\be
{\bar \rho}=\frac{\int d V T^{00}}{\int d V}\,.
\ee
Using Eq. (\ref{entropyeq1}), one may find the following evolution equation for the average energy density of domain walls
\be
{\dot {\bar \rho}} + H({\bar \rho} + {\cal {\bar P}}_\parallel)=0\,,
\label{entropyc1}
\ee
in a background with anisotropic expansion along the $x$- or $y$-direction. Since the distance between the domain walls would remain unchanged by an expansion parallel to the domain walls,  the average density of the domain wall network would remain a constant. Therefore we should have that ${\cal {\bar P}}_\parallel = -{\bar \rho}$. This is to be expected since we have previously demonstrated that $\rho \equiv T^{00}=-T^{xx}=-T^{yy}$. 

Consider now an anisotropic expansion along the $z$-direction only. In this case, it follows from entropy conservation that
\be
dU + {\cal {\bar P}}_\perp dV=0\,,
\label{entropyeq2}
\ee
where ${\cal {\bar P}}_\perp$ is the weighted volume average of $T^{zz}$,
\be
{\cal {\bar P}}_\perp=\frac{\int d V T^{zz}}{\int d V}\,.
\ee
In this situation, the distance between the domain walls would increase proportionally to the scale factor. Consequently, the average density of the domain wall network would evolve as ${\bar \rho} \propto  a^{-1}$, and thus ${\cal {\bar P}}_\perp=0$. This condition is consistent with the one derived in \cite{Avelino:2008ve} ($T^{zz}=0$), assuming energy-momentum conservation, for a single static planar domain wall.

\subsection{Linear momentum evolution}

Under a Lorentz boost with velocity $v$ along the $z$-direction  --- characterized by
\bq
\Lambda^{0'}_0&=&\Lambda^{z'}_z=\gamma \,, \quad \Lambda^{0'}_z=\Lambda^{z'}_0=\gamma v \,,\\
\Lambda^{y'}_y&=&\Lambda^{x'}_x=1\,,
\eq
---, the diagonal components of the energy-momentum tensor of a static planar domain wall oriented perpendicularly to the $z$-direction become
\bq
T^{0'0'}&=&\gamma^2T^{00}\,, \label{emdiag1}\\ 
T^{z'z'}&=&v^2 \gamma^2 T^{00}=v^2 T^{0'0'}  \,,\\
T^{x'x'}&=&T^{xx}=-T^{00}=(v^2-1)T^{0'0'}\,,\\
T^{y'y'}&=&T^{yy}=-T^{00}=(v^2-1)T^{0'0'}\label{emdiag4}\,.
\eq

Let us again consider a statistically homogeneous (but anisotropic) planar domain wall network with all domain walls oriented along the $z$-direction and assume that all domain walls move with the same speed $v$. Let us also assume an homogeneous and isotropic expansion (here we are implicitly neglecting the domain wall contribution to the cosmic energy budget). The first law of thermodynamics may, in this case, be written as

\be
dU+{\cal {\bar P}}_\parallel \frac{\partial V}{\partial x}dx+{\cal {\bar P}}_\parallel \frac{\partial V}{\partial y}dy+{\cal {\bar P}}_\perp \frac{\partial V}{\partial z}dz=0 \label{tflawdw}\,,
\ee
by separating the pressure contributions along each of the spatial directions. Eq. (\ref{tflawdw}) implies that

\be
{\dot {\bar \rho}} + H(3{\bar \rho} + 2{\cal {\bar P}}_\parallel + {\cal {\bar P}}_\perp)=0\,,
\label{entropyc2}
\ee
where ${\cal {\bar P}}_\parallel=(v^2-1) {\bar \rho}$ and ${\cal {\bar P}}_\perp=v^2 {\bar \rho}$ (as indicated by Eqs. (\ref{emdiag1})-(\ref{emdiag4})). Therefore, one has that
\be
{\dot {\bar \rho}}+H {\bar \rho}(1+3v^2)=0\,.
\label{rhoeqw}
\ee
If one assumes that the domain wall mass per unit area and the number of walls are conserved, the average domain wall energy density is given by
\be
{\bar \rho} \propto \gamma a^{-1}\,.
\label{densityb0}
\ee
Using Eqs. (\ref{rhoeqw}) and (\ref{densityb0}), one obtains the following evolution equation for the velocity of a domain wall
\be
{\dot v} + 3H(1-v^2)v=0\,,
\label{veqw}
\ee
which in turn implies that
\be
\gamma v \propto a^{-3}\,.
\ee
The linear-momentum of the domain walls in a fixed comoving volume is therefore conserved. Unlike in the case of the free particle, the evolution of the linear-momentum of planar domain walls is not a purely kinematic effect. The dynamical stretching of the domain walls due to the expansion of the universe causes a damping of their velocity that is significantly stronger than the (purely kinematic) damping experienced by point particles.

\subsection{\label{vos1} VOS model}

Similarly to the free particle case, the evolution equation of the RMS velocity may be obtained by averaging Eq. (\ref{veqw}). It takes the form
\be
{\dot {\bar v}} + 3H(1-{\bar v}^2){\bar v}=0\,.
\label{veqvosw}
\ee

The equation of state of a statistically homogeneous and isotropic domain wall network can be obtained by performing a Lorentz boost to the energy-momentum tensor of a single static planar domain wall and, then, averaging over domain wall speeds and all possible orientations of the domain walls \cite{kolb1994early}. The resulting equation-of-state parameter is given by
\be
w \equiv \frac{\cal {\bar P}}{\cal {\bar \rho}}=-\frac{2}{3}+{\bar v}^2\,.
\label{eqofstatewalls}
\ee
This equation has two important limits: the ultra-relativistic limit, with $w \sim 1/3$ --- in which domain walls behave effectively as radiation --- and the non-relativistic limit, with $w \sim -2/3$ --- in which the network has a negative average pressure. Note that by substituting Eq. (\ref{eqofstatewalls}) into Eq. (\ref{entropyc}), one would recover the weighted volume average of Eq. (\ref{rhoeqw}):
\be
{\dot {\bar \rho}}+H {\bar \rho}(1+3{\bar v}^2)=0\,.
\label{rhoeqwav}
\ee

For a statistically homogeneous and isotropic network, the characteristic length of the network is defined as
\be
{\bar \rho}=\frac{\sigma}{L}\,,
\label{Lwall}
\ee
where $\sigma$ is the domain wall mass per unit area. One may use Eqs. (\ref{rhoeqwav}) and (\ref{Lwall}) to obtain the following equation for the evolution of the characteristic lengthscale of a homogeneous and isotropic network of non-interacting planar domain walls
\be
{\dot L}-(1+3{\bar v}^2)HL=0\,.
\ee

Notice that, except for the difference in the numeric coefficients in the Hubble damping terms --- that results from the fact that domain walls are extended objects and thus stretched by expansion ---,  the VOS equations for point masses and domain walls have the same structure. 

\section{$p$-brane networks in $N+1$ dimensions \label{sec4}}

In this section, we apply the method used in the previous sections to the case of defects of arbitrary dimensionality $p$ in $N+1$-dimensional FLRW backgrounds. We then provide a derivation of the generalized VOS model for $p$-dimensional cosmic defect networks, including the variation of the defect mass per unit $p$-dimensional area as a new degree of freedom, and compute the corresponding scaling solutions.

\subsection{Energy-momentum tensor}

Consider a static curvatureless $p$-dimensional topological defect (or $p$-brane) in a $N+1$-dimensional homogeneous and isotropic spacetime and let us assume that it is oriented in such a way that the ${\hat i}=1,...,p$ spatial directions are parallel to the brane and the ${\tilde i}=p+1,...,N$ spatial directions are perpendicular to it (the $p$-brane is assumed to be planar along all the parallel ${\hat i}$ directions).  Let us also assume the defect is featureless, so that the physical velocity is perpendicular to it (i.e., $v$ only has components along the perpendicular ${\tilde i}$ directions). 

The energy-momentum tensor of the $p$-brane should be invariant under Lorentz boosts in the ${\hat j}$ directions parallel to the brane. Therefore, the only non-diagonal components of the energy-momentum tensor that can be non-vanishing are $T^{{\tilde i} {\tilde j}}$. On the other hand, the diagonal components should be such that
\be
\rho \equiv T^{00}=-T^{{\hat i}{\hat i}}\,.
\label{srhop}
\ee
In the particular case of point particles, we shall require that the $T^{0i}$ components of the energy-momentum tensor are equal to zero. Note however that this condition may not be a good approximation in the case of fast rotating masses. Note also that this requirement is automatically satisfied for $p\ge 1$.

Let us now consider a statistically homogeneous $p$-brane network and assume that the $p$-branes are static, non-interacting, and oriented along the same direction. As in the case of domain walls, the distance between the $p$-branes is unaffected by expansion along the parallel directions. This implies that
\be
{\cal {\bar P}}_\parallel \equiv \frac{\int d V T^{{\hat i}{\hat i}}}{\int d V}=-{\bar \rho}\,.\label{ppar}
\ee
On the other hand, expansion along the perpendicular directions dilutes the network. One then has that
\be
{\cal {\bar P}}_\perp \equiv \frac{\int d V T^{{\tilde i}{\tilde i}}}{\int d V}=0\,.
\ee
Note that Eq. (\ref{ppar}) is also a direct consequence of Eq. (\ref{srhop}).

\subsection{Linear momentum evolution}

If one performs a Lorentz  boost  with speed $v$ along the ${\tilde j}$ direction, the  diagonal components of the energy-momentum tensor of a static curvatureless $p$-brane transform as
\bq
T^{0'0'}&=&\gamma^2T^{00}\label{emdiagp1}\,, \\
T^{{\tilde j}'{\tilde j}'}&=&v^2 \gamma^2 T^{00}=v^2 T^{0'0'} \,,\\
T^{{\tilde i}'{\tilde i}'}&=&T^{{\tilde i}'{\tilde i}'}=0\\
T^{{\hat i}'{\hat i}'}&=&T^{{\hat i}{\hat i}}=(v^2-1) T^{0'0'}\,\label{emdiagp2} \,.
\eq
for $\tilde{i} \neq \tilde{j}$.

Let us again consider a homogeneous topological defect network in an homogeneous and isotropic expanding background, and assume that all the defects have the same velocity along a fixed direction $x^{\tilde j}$. Again, we are implicitly neglecting the $p$-brane contribution to the cosmic energy budget. Note that, as indicated by Eqs. (\ref{emdiagp1})-(\ref{emdiagp2})), the only diagonal pressure component of the energy-momentum tensor along the  perpendicular directions is that along the velocity direction $x^{\tilde j}$. All other components vanish. 

Let us define ${\cal {\bar P}}_{\perp*}=T^{{\tilde j}{\tilde j}}$ and $dx_*=dx^{\tilde j}$. It follows from the first law of thermodynamics and from entropy conservation in an adiabatically expanding background that 
\be
dU+\left(p{\cal {\bar P}}_\parallel\frac{\partial V}{\partial x^{\hat i}}dx^{\hat i}+{\cal {\bar P}}_{\perp*}\frac{\partial V}{\partial x^*}dx^*\right)=0\,,
\label{entropycg}
\ee
with $U={\bar \rho V}$ and $V \propto a^N$, and where ${\cal {\bar P}}_\parallel=(v^2-1)\bar{\rho}$ and ${\cal {\bar P}}_{\perp*}=v^2\bar{\rho}$ (see Eqs. (\ref{emdiagp1})-(\ref{emdiagp2})). Hence, one has that
\be
{\dot {\bar \rho}}+ H(N{\bar \rho} + p{\cal {\bar P}}_\parallel + {\cal {\bar P}}_{\perp*})=0\,,
\label{rhocg}
\ee
or, equivalently,
\be
\dot{\bar \rho}+H\bar{\rho}\left[N-p+\left(p+1\right)v^2\right]=0\,.
\ee
Taking into account that the average $p$-brane energy density scales as
\be
{\bar \rho} \propto \frac{\sigma_p \gamma}{a^D}\,,
\label{densitypb}
\ee
(where $D=N-p$ is the brane co-dimension) and substituting Eq. (\ref{densitypb}) into Eq. (\ref{rhocg}), one obtains 
\be
{\dot v} + (p+1)H(1-v^2)v=0\,.
\label{veqpb}
\ee
Therefore, the linear-momentum of the $p$-branes in a fixed comoving volume is conserved:
\be
\gamma v \propto a^{-(p+1)}\,.
\ee

Let us now relax the assumption that all the defects have the same velocity. The characteristic lengthscale of a $p$-brane network $L$ is usually defined as
\be
{\bar \rho}=\frac{\sigma_p}{L^D}\,,
\label{Ldef}
\ee
where $\sigma_p$ is the defect mass per unit $p$-dimensional area. The evolution equation for the characteristic lengthscale of an homogeneous and isotropic network of non-interacting $p$-branes may then be written as
\be
{\dot L}-HL-\frac{L}{D \ell_d}{\bar v}^2 =0\,,
\label{Lvosw}
\ee
where $\ell_d^{-1}=(p+1)H$ is the damping lengthscale. Following the procedure described in the previous sections, one may find the following evolution equation for the RMS velocity of the $p$-brane network
\be
{\dot {\bar v}} + (1-{\bar v}^2)\frac{\bar v}{\ell_d}=0\,,
\label{vvosw}
\ee
The frictional force caused by the interaction of the topological defects with ultrarelativistic particles or other frictional sources --- characterized by a homogeneous friction lengthscale $\ell_f$ --- may be included in the VOS equations of motion (Eqs (\ref{Lvosw}) and (\ref{vvosw})) by redefining the damping lengthscale as $\ell_d^{-1}=(p+1)H+\ell_f^{-1}$.

\subsection{Topological defects with varying $\sigma_p$}

Here, we shall consider a new dynamical degree of freedom in the evolution of defect networks: $\sigma_p$. Although the value of $\sigma_p$ is usually assumed to be a constant when considering a specific topological defect network, its evolution may be coupled to that of a background scalar field $\phi$. Here, we shall assume the scalar field $\phi$ to be homogeneous, thus neglecting fifth forces associated to the gradients of the field and small-scale backreaction effects \cite{Avelino:2015fka}. In the absence of friction,  conservation of the linear momentum of a curvatureless $p$-brane with mass per unit $p$-dimensional area $\sigma_p(\phi)$ implies that
\be
\sigma_p \gamma v \propto a^{-(p+1)}\,.
\ee
 The equation of motion for the velocity of the $p$-brane is thus of the form
\be
{\dot v} + (1-v^2)\frac{v}{\ell_d}=0\,,
\ee
with a modified damping scale 
\be
\ell_d^{-1}=(p+1+\chi)H\,,
\label{elld}
\ee
where we have introduced $\chi={\dot \sigma}_p/(\sigma_p H)$. The effect of friction may also be included in the definition of $\chi$ by rewriting it as
\be
\chi=\frac{1}{H}\left(\frac{\dot \sigma_p}{\sigma_p}+\ell_f^{-1}\right)\,.
\label{chi}
\ee
By averaging over all defect orientations and speeds, and assuming statistical homogeneity and isotropy, one recovers the macroscopic equation for the evolution of the RMS velocity (Eq. (\ref{vvosw})), but now with $\ell_d$ given by Eqs. (\ref{elld}) and (\ref{chi}).

In order to compute an evolution equation for ${\bar \rho}$, let us again consider an homogeneous and isotropic network of non-interacting curvatureless $p$-branes, wherein all branes have the same speed $v$. In this case, the average density is given by Eq. (\ref{densitypb}) and, consequently,
\be
\frac{\dot {\bar \rho}}{\bar \rho}-\frac{\dot \sigma_p}{\sigma_p}=\frac{\dot \gamma}{\gamma}-D\frac{\dot a}{a}=-\frac{v^2}{\ell_d}-DH\,.
\ee

Relaxing the assumption that all the defects have the same velocity, so that each defect has its own arbitrary velocity, one recovers Eq. (\ref{Lvosw}), with $\ell_d$ given by Eqs. (\ref{elld}) and (\ref{chi}) (after averaging over all speeds and using the definition of $L$ in Eq. (\ref{Ldef})). Note that the energy transferred from the scalar field to the $p$-brane network per unit of time is given by
\be
\left(\frac{\dot {\bar \rho}}{\bar \rho}\right)_\phi=\frac{\dot \sigma_p}{\sigma_p}(1-{\bar v}^2)\,.
\ee

\subsection{Generalized VOS model for $p$-brane network evolution}

According to the generalized VOS model for $p$-branes, the cosmological evolution of a statistically homogeneous and isotropic $p$-brane network is described by the following equations
\bq
\frac{d{\bar v}}{dt}&+&\left(1-{\bar v}^2\right)\frac{{\bar v}}{\ell_d}=\left(1-{\bar v}^2\right)\frac{k}{L}\,, \label{VOS_v}\\
\frac{dL}{dt}&-&HL-\frac{L}{D \ell_d}{\bar v}^2=\frac{{\tilde c}}{D}{\bar v}\,,
\label{vos-L}
\eq
where $k({\bar v})$ and ${\tilde c}$ are the curvature and (phenomenological) energy loss parameters, respectively. The curvature parameter $k({\bar v})$, rigorously defined in \cite{Avelino:2011ev,Sousa:2011ew}, describes the conversion of rest mass energy into kinetic energy (and vice-versa) by the network, thus describing the acceleration of the $p$-branes due to their curvature. Note that Eqs. (\ref{VOS_v}) and (\ref{vos-L}) are identical to Eqs.  (\ref{vvosw}) and (\ref{Lvosw}), apart from the curvature term on the right hand side of Eq. (\ref{VOS_v}), and the energy loss term associated with $p$-brane reconnection on the right hand side of Eq. (\ref{vos-L}). Therefore, the simple derivation based on thermodynamic considerations we have considered here captures the essential aspects driving topological defect dynamics.

\subsection{Scaling Regimes}

The scaling regimes of $p$-brane networks in FLRW backgrounds with constant $\sigma_p$ have been studied in detail in Refs. \cite{Avelino:2011ev,Sousa:2011ew}. Note however that, if one allows for a time variation of $\sigma_p$, the characteristics of these regimes may be altered and their existence may depend on the specific evolution of $\chi$ as a function of time. Here, we shall revisit these regimes to determine the effect that allowing for the coupling of $\sigma_p$ to a background scalar field has on them.

Let us start by assuming that frictional forces are negligible (i.e., $\ell_f=+\infty$) and by studying the different scaling regimes that may arise throughout the cosmological evolution of frictionless networks. In a FLRW universe with a decelerating power-law expansion --- $a \propto t^\beta$ with $0\le \beta<1$ --- Eqs. (\ref{VOS_v}) and (\ref{vos-L}) are known to admit linear attractor solutions of the form
\be
L=\xi t \qquad \mbox{and} \qquad \bar v=\mbox{constant}\,,
\label{linearscalinga}
\ee
for $\chi=0$. If one allows for a time variation of $\sigma_p$, solutions of this kind are only attainable if $\chi$ is constant, or, equivalently, if $\sigma_p$ is of the form $\sigma_p \propto a^{\chi}$. In this case, the linear scaling regime would be characterized by
\bq
\xi &=& \sqrt{\left|\frac{k(k+{\tilde c})}{\beta (1-\beta)D(p+1+\chi)}\right|}\label{linearscaling}\,,\\
{\bar v} &=& \sqrt{\frac{(1-\beta)kD}{\beta(k+{\tilde c})(p+1+\chi)}}\label{linearscaling1}\,.
\eq
The effect of allowing for variations of $\sigma_p$ of this form on the value of the scaling coefficients is significant and varied. For positive values, $\chi$ hinders defect growth, leading to a larger number density of defects. Note however that this effect does not seem to be dramatic enough to frustrate the networks for reasonable values of $\chi$. For negative values of $\chi$, variations of $\sigma_p$ create an additional source of acceleration of the branes, leading to the dilution of the network. This scaling solution is not attainable for all values of $\chi$. Defect growth is necessarily constrained by causality and, therefore, the characteristic lengthscale should be smaller than the particle horizon at any given time. Moreover, the RMS velocity must be smaller than unity ($\bar{v}<1$). These conditions prevent the attainment of this regime for large negative values of $\chi$. Note, however, that a detailed study of the effect of time variations of $\sigma_p$ on the mean curvature and on the efficiency of the energy-loss mechanism would be necessary to precisely quantify the minimum value of $\chi$.
 
If initially the characteristic lengthscale is sufficiently large (with $HL \ll {\bar v}^2 L/(D\ell_d)$ and $HL\ll {\tilde c}{\bar v}/D$), the network may experience a transient regime  during which it is conformally stretched with

\be
L \propto a\,.
\label{stretch}
\ee
However, for a decelerating background expansion, $HL$ is necessarily a decreasing function of time. Therefore, the energy loss caused by interactions between the defects eventually plays a significant role in the dynamics, bringing the regime to an end. Note however that this is no longer necessarily true if the universe enters an inflationary stage, with $\ddot{a}>0$.

Other scaling solutions are also possible. For instance, if the $p$-branes are non-relativistic and ${\dot {\bar v}} \ll {\bar v}/\ell_d$, the characteristic velocity of the network is given by
\be
{\bar v} \propto \frac{\ell_d}{L}\,.
\ee
In this case, if $\ell_d={\rm const}$, the RMS velocity and the characteristic length of the network will be inversely proportional. This regime is particularly important in various condensed matter and biological systems wherein the network dynamics is curvature driven and the background expansion can be neglected \cite{Avelino:2010qf,2012PhRvE..86c1119A}.

Let us now consider the case in which friction plays a relevant role in network dynamics. If the network density is initially low, its characteristic lengthscale would be such that $L\ll \ell_f$ and, as a consequence, the network experiences a stretching regime of the form of Eq. (\ref{stretch}), during which ${\bar v}=\ell_f/L$. However, as the network evolves, it starts to experience a considerable energy loss due to self-interaction  $HL \sim {\tilde c}{\bar v}/D$. As a result, $p$-brane network experiences a different scaling regime --- the Kibble regime --- characterized by

\be
L\propto \sqrt{\frac{\ell_f}{\left|H\right|}}\,,\qquad\mbox{and}\qquad {\bar v}\propto \sqrt{\ell_f\left|H\right|}\,.
\ee
Note that if, at the moment of formation of the branes, the density of the network is high enough, the Kibble regime may occur immediately.

The evolution of both ${\bar v}$ and $L$ in these friction-dominated regimes is also sensitive to the variation of $\sigma_p$. The interaction of $p$-branes with the ultrarelativistic particles in a radiation fluid with density $\rho_{\rm rad}$ results in a friction length $\ell_f \propto \sigma_p \lambda^{p+1-N}/\rho_{\rm rad}$, where $\lambda$ is the typical wavelength of the particles. Therefore, if $\sigma_p$ varies with time, the evolution of $\ell_f$ --- which would otherwise scale as $\ell_f\propto a^{p+2}$ --- is linearly affected.

Therefore, the evolution of $L$ and $\bar v$ throughout cosmological history is sensitive to time variations of $\sigma_p$. Finally, even for a fixed evolution of ${\bar v}$ and $L$ the average topological defect energy density ${\bar \rho}$ would vary proportionally $\sigma_p$. Hence, a variation of $\sigma_p$ with cosmic time is expected to have a significant impact on the observational signatures of cosmic defects, independently of their dimensionality.

\section{\label{conc} Conclusions}

In this paper, we have shown that the equation of state and some of the essential aspects of the dynamics of cosmic defects of arbitrary dimensionality are determined by simple thermodynamic conditions. We have taken advantage of this fact to derive the dependence of the VOS model for the macroscopic dynamics of topological defect networks in homogeneous and isotropic universes on the dimensionality of the defects and on the number of space-time dimensions. We incorporated a parameterization of  the changes to the VOS equations of motion associated to the interaction of the defects with other fields, including, in particular, a new dynamical degree of freedom associated to the variation of the mass per unit $p$-area of the defects. This approach provides a more direct and illuminating derivation of the VOS equations of motion describing the evolution of the RMS velocity and the characteristic (macroscopic) length of the network. Moreover, it may be useful for pedagogical purposes as a fairly intuitive way of addressing the complex subject of defect evolution.

We have shown that both frictional effects as well the new dynamical degree of freedom associated to the variation of the defect mass per unit $p$-area may significantly affect the network dynamics and the corresponding scaling solutions. Apart from the changes to the VOS equations of motion (which can be incorporated in a redefinition of the damping scale), a variation of the defect mass per unit $p$-dimensional area has a direct impact on the evolution of the network energy density.  Therefore, such a variation, if it occurs, is expected to leave observational imprints, for instance, in the shape of the power spectra of the cosmic microwave and stochastic gravitational wave backgrounds generated by cosmic defects.

\begin{acknowledgments}

P.P.A. is supported by Funda{\c c}\~ao para a Ci\^encia e a Tecnologia (FCT) through the Investigador FCT contract reference IF/00863/2012 and POPH/FSE (EC) by FEDER funding through the program Programa Operacional de Factores de Competitividade - COMPETE. L.S. is supported by FCT and and POPH/FSE through the grant SFRH/BPD/76324/2011. Funding of this work was also provided by the FCT grant UID/FIS/04434/2013.

\end{acknowledgments}


\bibliography{VOSTH}

\end{document}